\documentclass[pdflatex,sn-mathphys-num]{sn-jnl}

\usepackage{graphicx}%
\usepackage{multirow}%
\usepackage{amsmath,amssymb,amsfonts}%
\usepackage{amsthm}%
\usepackage{mathrsfs}%
\usepackage[title]{appendix}%
\usepackage{xcolor}%
\usepackage{textcomp}%
\usepackage{manyfoot}%
\usepackage{booktabs}%
\usepackage{algorithm}%
\usepackage{algorithmicx}%
\usepackage{algpseudocode}%
\usepackage{listings}%
\usepackage{siunitx}
\usepackage{graphicx}
\usepackage{multirow}
\DeclareSIUnit\gauss{G}

\usepackage{caption}
\usepackage{subcaption}
\usepackage{graphicx}

\theoremstyle{thmstyleone}%
\theoremstyle{thmstyletwo}%

\theoremstyle{thmstylethree}%

\begin{document}

\title[Article Title]{An IMU Dataset for Human Activity Recognition to Support Independent Living in Smart Homes (IMU-HAR-IL)}


\author*[1,2]{\fnm{Moid} \sur{Sandhu}}\email{moid.sandhu@csiro.au}

\author[1,3]{\fnm{Wei} \sur{Lu}}\email{wei.lu@csiro.au}

\author[4]{\fnm{Brano} \sur{Kusy}}\email{brano.kusy@csiro.au}

\author[1,5]{\fnm{David} \sur{Silvera-Tawil}}\email{david.silvera-tawil@csiro.au}

\affil[1]{\orgname{Commonwealth Scientific and Industrial Research Organisation (CSIRO)}, \orgaddress{\street{Herston}, \city{Brisbane}, \postcode{4029}, \state{QLD}, \country{Australia}}}

\affil[2]{\orgdiv{School of Computer Science}, \orgname{Queensland University of Technology}, \orgaddress{\city{Brisbane}, \postcode{4000}, \state{QLD}, \country{Australia}}}

\affil[3]{\orgdiv{School of Electrical Engineering and Computer Science}, \orgname{The University of Queensland}, \orgaddress{\city{Brisbane}, \postcode{4072}, \state{QLD}, \country{Australia}}}

\affil[4]{\orgname{Commonwealth Scientific and Industrial Research Organisation (CSIRO)}, \orgaddress{\street{Pullenvale}, \city{Brisbane}, \postcode{4069}, \state{QLD}, \country{Australia}}}

\affil[5]{\orgname{International Centre for Future Health Systems, University of New South Wales (UNSW)}, \orgaddress{\city{Sydney}, \postcode{2052}, \state{NSW}, \country{Australia}}}


\abstract{
This document introduces HAR-IMU-IL, a dataset developed for human activity recognition (HAR) using inertial measurement unit (IMU) sensors within a smart home environment with a focus to support objective functional assessment of older adults' independent living (IL). In particular, HAR-IMU-IL includes recordings of 50 participants performing 17 clinically relevant activities of daily living, spanning 4 functional domains essential for independent living: mobility, hygiene, nutrition and hydration, and medication intake. The dataset was collected using 30 IMU sensors, comprising both wearable and object-mounted devices integrated within a real-world residential setting. The dataset includes multi-sensor inertial data captured under realistic, unconstrained conditions, together with detailed annotations ensuring high temporal accuracy and consistency across sensors. A comprehensive data collection protocol was implemented to preserve ecological validity and enable reliable multi-sensor synchronisation. HAR-IMU-IL provides a large-scale, functionally grounded resource for advancing and benchmarking machine learning and artificial intelligence approaches for HAR in home settings. We further demonstrate its utility by developing models capable of accurately recognising both activities and broader functional domains using wearable and object-mounted sensors. These capabilities highlight the dataset’s potential to enable applications in continuous activity monitoring, functional health assessment, smart home automation, and assistive technologies that support independent living.
}

\maketitle

\section{Background \& Summary }\label{sec1}
    The rapid advancement and widespread adoption of internet of things (IoT) sensing systems have paved the way for increasingly sophisticated human activity recognition (HAR) mechanisms~\cite{sandhu2023self, bouchabou2021survey}. These IoT sensors play a crucial role in supporting healthcare, fitness, and activity monitoring, as well as  optimising home automation in everyday living environments \cite{mahato2024hybrid, torres2025smart}. By leveraging a range of sensing technologies, HAR systems facilitate continuous monitoring of daily activities, providing valuable insights into an individual's well-being, enabling the detection of anomalies, and supporting the assessment of independent living in real‑world settings \cite{sandhu2025fusing, baig2019systematic, alaghbari2022activities}. Such capabilities offer significant potential to enable timely and targeted interventions when needed \cite{sandhu2025feasibility, oyibo2023using, lu2025impact}. This is particularly beneficial for older adults and individuals living with disabilities, as it facilitates proactive care and personalised assistance while promoting autonomy and safety at home~\cite{lu2025impact}.

    A range of sensing modalities have been explored for HAR, each offering distinct advantages and limitations in home and smart living environments~\cite{sandhu2024internet}. These modalities include vision-based systems (e.g., cameras)~\cite{fleck2008smart}, radio frequency (RF)-based methods (e.g., Wi-Fi, radar, and RFID)~\cite{usman2022intelligent}, and motion sensors (e.g., accelerometers, magnetometers, and gyroscopes)~\cite{milosevic2020kinect}. Vision-based systems capture rich contextual information about daily activities; however, they raise significant privacy concerns and typically require substantial computational resources for data processing which can limit their suitability for long‑term in‑home deployment~\cite{sandhu2023self}. RF-based methods provide contactless sensing and can detect movement through obstacles such as walls; however, they are vulnerable to signal interference, operate in increasingly congested spectral bands, often require high bandwidth, and impose substantial computational demands~\cite{uysal2022new, chen2021rf}.

    Among these modalities, motion sensors have emerged as a particularly suitable choice for HAR in home environments due to their favourable balance between privacy, accuracy, and ease of integration \cite{sandhu2025fusing}. Unlike video-based approaches, motion sensors do not capture audio-visual data, making them inherently privacy-preserving and well suited to domestic settings. Additionally, they are cost-effective, energy-efficient, and can be seamlessly embedded within both wearable devices and every day household objects as part of a smart home infrastructure~\cite{sandhu2024internet}. Their ability to support continuous and unobtrusive monitoring makes motion sensors well suited for real-world applications such as personal health tracking, physical activity monitoring, and supporting independent living~\cite{bolam2021remote, csengul2022deep, sandhu2023self}.

    Data collected from motion sensors is widely used to develop machine learning models for real-time activity recognition. However, a key challenge in advancing accurate and reliable home-based HAR models is the limited availability of comprehensive datasets. Many existing studies rely on a datasets with small (n$<$15) participant cohorts and a restricted set of activities~\cite{roggen2010collecting}, often overlooking key activities that are clinically relevant for assessing the functional independence of older adults~\cite{shahid2022detecting, sokullu2020iot, vanus2017monitoring, ramos2022sdhar}. Moreover, a large proportion of published datasets are collected in structured or laboratory settings, limiting their generalisability to real-world home environments~\cite{roggen2010collecting, sikder2021ku, stisen2015smart, alshammari2018simadl}. Finally, prior work has often focused on either wearable or object-mounted sensors in isolation, rather than leveraging their complementary strengths through sensor fusion to maximise activity recognition performance~\cite{sikder2021ku, stisen2015smart, lago2017contextact, alshammari2018simadl}.

    Collecting comprehensive real-world activity data is critical for the development of robust and generalisable machine learning models, particularly when focusing on real-world activities that are relevant to independent living assessment and support. Moreover, leveraging data collected in real-world settings can enhance the ecological validity of these activity recognition models, facilitating their deployment into practical, home-based monitoring systems and ultimately contributing to safer and more sustainable ageing in place.

    \begin{table}[b!] 
    \caption{Clinical assessment tools reviewed to inform activity selection for functional assessment and independent living evaluation.}\label{table:assessment_tools}
    \centering
    \begin{tabular}{p{0.65\linewidth} p{0.25\linewidth}}
    \toprule
    \textbf{Tools}	& \textbf{Functional focus} \\
    \midrule
    Barthel Index/Modified Barthel Index~\cite{wang2023comparison} & \multirow{5}{2.8cm}{Functional independence and abilities} \\
    Katz Activities of Daily Living~\cite{wallace2007katz} \\
    Functional Independence Measure~\cite{kidd1995functional} \\
    Older Americans Resources and Services Program~\cite{pfeiffer1975older} \\
    Lawton Brody Instrumental Activities of Daily Living~\cite{fish2011lawton} \\
    \hline
    FRAIL Scale~\cite{woo2015frailty} & \multirow{4}{2.8cm}{Frailty} \\
    Rockwood Clinical Frailty Scale~\cite{church2020scoping} \\
    Fried Frailty Scale~\cite{bahat2021success} \\
    Edmonton Frail Scale~\cite{rolfson2006validity} \\
    \hline
    De Moreton Mobility Index~\cite{de2008morton} & \multirow{2}{2.8cm}{Functional mobility} \\
    Short Physical Performance Battery~\cite{pavasini2016short} \\
    \hline
    Mini-Nutritional Assessment~\cite{MNA:2025:Online} & Nutrition \\
    \bottomrule
    \end{tabular}
    \end{table}

    This paper presents HAR-IMU-IL, a novel dataset for HAR in home environments using wearable and object-mounted IMU sensors. In contrast to existing work~\cite{shahid2022detecting, sokullu2020iot, vanus2017monitoring, ramos2022sdhar}, the dataset is explicitly grounded in functional assessment principles, incorporating standard clinical frameworks and expert clinical input to define 17 critical activities across 4 functional areas domains directly relevant to evaluating independent living capabilities. To enable clinically meaningful and ecologically valid data collection, a smart home environment was instrumented with a dense deployment of 30 wearable and object-mounted IMU sensors, and data were collected from 50 adult participants performing everyday activities under naturalistic conditions. The resulting dataset can be employed to develop and evaluate various machine learning models for real-time HAR, particularly for functional monitoring and independent living support. In addition, this work establishes baseline performance benchmarks that provide a foundation for future HAR model development and evaluation.

\section{Methods}\label{sec2}

\subsection{Activity Selection}

\begin{table}[b!] 
\caption{List of functional areas and key activities performed by participants during the data collection.}\label{table:activities_list}
\centering
\begin{tabular}{p{0.20\linewidth} p{0.70\linewidth}}
\toprule
\textbf{Functional area} & \textbf{Human activities} \\
\midrule
Mobility &	Sitting, standing, walking, lying, stairs, sit-stand transitions, sit-lay transitions.\\
\hline
Hygiene & Washing face, brushing teeth, using toilet, showering, dressing.
\\
\hline
Nutrition \& hydration & Preparing meal, drinking, eating, empty kitchen bin.
\\
\hline
Medication & Taking medicine. \\

\bottomrule
\end{tabular}
\end{table}

\begin{figure}[b!]
\centering
\includegraphics[width=0.7\linewidth]{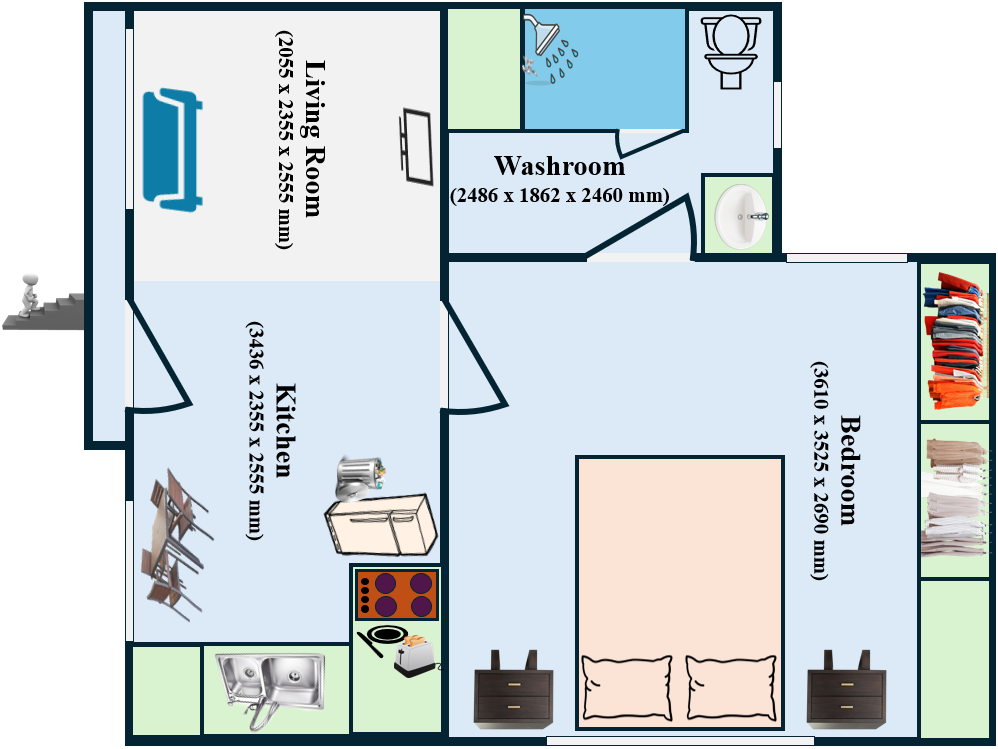}
\caption{Layout of the home environment.}
\label{fig:smart_home_layout}
\end{figure}

\begin{table}[b!] 
\caption{Description and sequential order of activities performed by participants during the data collection.}\label{table:activity_interpretation}
\centering
\begin{tabular}{p{0.23\linewidth} p{0.80\linewidth}}
\toprule
\textbf{Activity} 	& \textbf{Description}\\
\midrule
Sitting & Maintaining a seated posture on a chair. \\
Sit-stand transitions 	& Alternating between standing on the floor and sitting on a chair in a repeated sequence. \\
Standing & Maintaining an upright standing posture. \\
Stairs & Ascending and descending wooden stairs located outside the apartment. \\
Walking & Ambulating within an indoor environment between living room and bedroom. \\
Dressing & Opening the bedroom cupboard, removing a hanger, changing the jumper, returning the hanger and clothing to the cupboard, and closing the cupboard door. Some participants simulated the dressing activity by not removing the jumper and instead returning an empty hanger to the cupboard. \\
Lying 	& Maintaining a lying posture on a bed. \\
Sit-lay transitions & Alternating between sitting on a bed and lying on a bed in a repeated sequence. \\
Taking medicine & Retrieving medication from a box inside a bedside drawer and simulating pill intake using a glass of water.\\
Brushing teeth  & Putting toothpaste on a toothbrush and simulating the brushing close to basin in the bathroom. \\
Washing face & Simulating this activity of washing the face using a face wash liquid at a bathroom basin. \\
Using toilet & Opening the lid, sitting on a toilet seat and simulating toilet use, closing the lid. \\
Showering & Opening the shower tap, simulating body washing, closing the shower tap. \\
Dressing & Opening the bedroom cupboard, retrieving a hanger with clothes, changing into a jumper, returning the empty hanger to the cupboard, and closing the cupboard door. For some participants, the jumper-changing step was simulated rather than physically performed. \\
Preparing meal & Preparing a simple meal by retrieving bread from a refrigerator and spread/jam from a kitchen pantry, followed by using a toaster. \\
Eating & Consumption of the prepared meal (toast with spread) while seated at a table and chair. Some of the participants simulated the meal consumption step.\\ 
Drinking & Drinking water using a glass filled from a kitchen tap. Some of the participants simulated the drinking step.  \\
Empty kitchen bin & Removing the kitchen bin liner and placing it outside the main entrance door. Some of the participants simulated the removal of kitchen bin liner. \\
\bottomrule
\end{tabular}
\end{table}

    A set of 17 key activities of daily living was identified based on a comprehensive review of 12 standard clinical assessment tools --- summarised in Table~\ref{table:assessment_tools} --- and extensive consultations with six clinicians with experience in aged care and disability, including two physiotherapists, two occupational therapists, a dietitian, and a music therapist. These activities span four critical functional domains: mobility, hygiene, nutrition and hydration, and medication intake (Table~\ref{table:activities_list}), and serve as clinically relevant indicators for evaluating the independent living capabilities of older adults. Their selection was guided by two primary considerations: their relevance to established clinical frameworks for functional assessment and independent living evaluation, and their feasibility for reliable detection using cost-effective, low-power IMU sensors. Although the inclusion of additional activities could further enrich independent living assessment, the scope was deliberately limited to 17 activities to balance clinical relevance with the practical constrains of data collection and the overall study manageability.

\subsection{Setup} 

    Data collection was conducted in a fully furnished one‑bedroom apartment in Brisbane, Australia, with the layout illustrated in Fig.~\ref{fig:smart_home_layout}. Healthy adult participants were recruited to perform the selected activities of daily living (Table~\ref{table:activities_list}) under realistic, naturalistic conditions. Table \ref{table:activity_interpretation} provides a detailed description of each activity and illustrates the chronological order of activities carried out by the participants throughout the experimental session. 

    A total of 30 Xsens DOT IMU sensors were used for data acquisition~\cite{movella_dot}. Out of these, 11 sensors were strategically placed on the participant’s body to capture fine-grained motion dynamics associated with daily activities, while the remaining 19 sensors were deployed on commonly interacted objects across different rooms in the smart home environment, as detailed in Table~\ref{tab1e:sensor_placement}. This hybrid configuration enabled the simultaneous capture of both human-centric motion patterns and object-level interaction signals, supporting robust activity recognition in realistic settings.

\begin{figure}[b!]
    \centering
    \includegraphics[width=\linewidth]{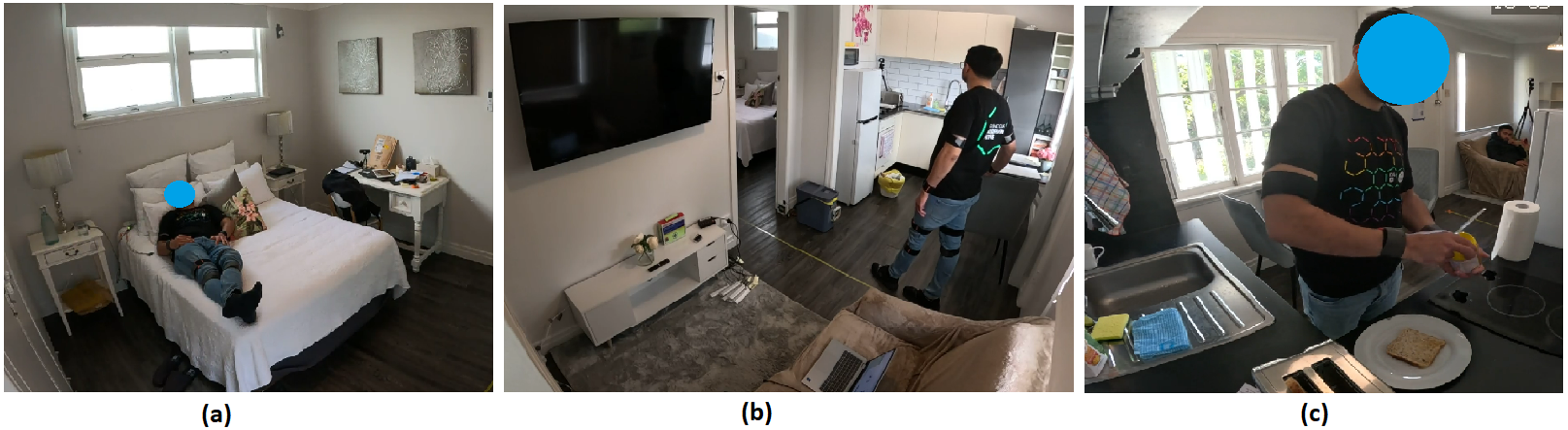}
    \caption{Data collection scenario in (a) bedroom, (b) living room, and (c) kitchen.}
    \label{fig:video_pics}
    \end{figure}

    In addition to the IMU-based sensing infrastructure, a multi-view video recording setup was employed with three cameras installed in the bedroom, living room, and kitchen. These cameras were used solely for ground-truth annotation and temporal activity labelling, ensuring accurate synchronisation and validation of the sensor-based data streams. Representative snapshots from the video recording setup are illustrated in Fig.~\ref{fig:video_pics}.

    \begin{table}[b!] 
    \caption{Placement of IMU sensors on the human body and selected objects in the home environment.}\label{tab1e:sensor_placement}
    \centering
    \begin{tabular}{p{0.15\linewidth} p{0.70\linewidth}}
    \toprule
    \textbf{Place}	& \textbf{Detailed location} \\
    \midrule
    \raggedright \multirow{3}{2.8cm}{Human body\\ (11 sensors)} &	Left wrist (LW), right wrist (RW), left upper arm (LUA), right upper arm (RUA), waist (WT), left thigh (LTH), right thigh (RTH), left leg (LLG), right leg (RLG), left shoe (LSH), right shoe (RSH).
    \\ \hline
    \multirow{5}{1.8cm}{Objects\\ (19 sensors)} & \textbf{Kitchen}: Table, chair, water tap, jam bottle, toaster, glass, plate, knife, kitchen bin. \\ 
     & \textbf{Bedroom}: Bed, medication box, glass, clothes hanger (in wardrobe).
    \\  
    & \textbf{Bathroom}: Water tap, facewash bottle, toilet lid, shower tap, toothbrush, toothpaste.
    \\
    \bottomrule
    \end{tabular}
    \end{table}

    Necessary precautions were taken to prioritise participant safety during data collection. A first-aid kit was kept readily available, and a trained assistant was designated to respond promptly in the event of any medical emergencies. These measures were implemented to ensure that participants could engage in the study with confidence, knowing that their well-being was of paramount importance.

\subsection{Procedure}

    Prior to participation, each participant received a comprehensive briefing detailing the purpose of the study, the data collection procedures, and the intended use of the collected data. This briefing ensured that participants clearly understood their role in the study and the broader implications of their contribution. Participants were then provided with an informed consent form detailing the study’s procedures, potential risks, and ethical considerations. Additionally, a withdrawal form was supplied describing the process for requesting the removal of their activity data should they wish to do so at any stage of the study.

    To uphold ethical standards and ensure voluntary participation, individuals were explicitly informed of their right to decline any activity from the predefined list or to withdraw from the study at any time without providing justification. Participants were also instructed to promptly report any discomfort, fatigue, or injury experienced during or immediately after the data collection session. Acknowledgment of these conditions was confirmed through the participants' signatures, names, and dates on the consent form before they were formally enrolled in the study.

    \begin{figure}[b!]
    \centering
    \includegraphics[width=\linewidth]{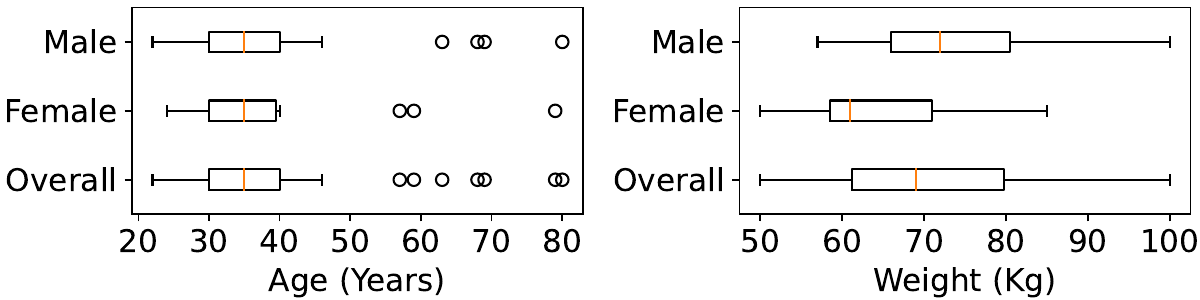}
    \caption{Demographics of participants.}
    \label{fig:Demographics_participants}
    \end{figure}
    
    In total, data were collected from 50 participants (35 male and 15 female) with an average age of 38.74 years and an average weight of \SI{71.12}{\kilo\gram}, as summarised in Fig.~\ref{fig:Demographics_participants}. Each participant performed multiple repetitions (up to four) of each activity according to the predetermined sequence presented in Table \ref{table:activity_interpretation}. The protocol allowed flexibility in the number of repetitions completed, enabling participants to stop before reaching four repetitions based on their comfort level and willingness to continue. As a result, the number of repetitions varied across participants.
    Several activities, including medication intake, brushing teeth, washing face, using toilet, and showering, were partially simulated by all participants. In contrast, activities such as dressing, meal preparation, eating, and drinking were only partially simulated by a subset of participants; while some participants performed these activities naturally, others enacted simulated versions of the same tasks. Table \ref{table:activity_interpretation} provides a detailed overview of the specific aspects of each activity that were simulated in this study. To minimise fatigue and ensure consistency in task performance, short rest periods of 5-15 minutes were provided between consecutive repetitions.

    During each session, a project team member manually annotated activity labels in real time as participants performed the prescribed activities. To ensure the accuracy and consistency of these annotations, video recordings were concurrently captured (Fig.~\ref{fig:video_pics}). These recordings were used exclusively for post‑hoc validation of the activity labels and were not included in the released dataset, ensuring participant privacy and compliance with ethical data‑handling practices.

    Ethics approval to complete this study was obtained from CSIRO's Health and Medical Human Research Ethics Committee, before conducting this study (Approval number: 2024\_018\_LR). All participants provided informed written consent before participating in the study.

\subsection{Data Acquisition}
\begin{figure*}[b!]
    \centering
    \includegraphics[width=\linewidth]{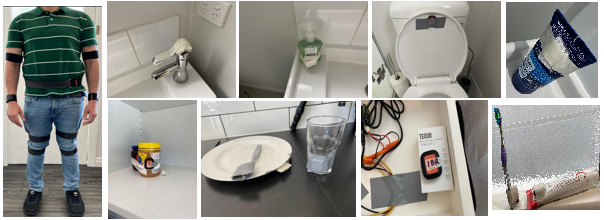}
    \caption{Examples of IMU sensor deployments on various objects and on the human body.}
    \label{fig:xsens_sensor_deployment}
    \end{figure*}

    Data was collected using 30 Xsens DOT sensors using the in-built recording mode~\cite{movella_dot}. These are light-weight IMU sensors (\SI{11.2}{\gram}) widely used for precision context tracking. In total, 11 sensors were attached to various parts of the human body as wearable devices and 19 sensors were attached to different objects within the home environment (Fig.~\ref{fig:xsens_sensor_deployment} and Table \ref{tab1e:sensor_placement}). These sensors were configured to collect motion data from a 3-axis gyroscope: \SI{\pm2000}{\degree/\second}, 3-axis accelerometer: \SI{\pm16}{\gram}, and 3-axis magnetometer: \SI{\pm8}{\gauss} at a sampling rate of \SI{60}{\hertz}.

    To facilitate reliable data synchronisation, the motion sensors were organised into three groups (Table \ref{tab1e:sensor_placement}): wearable sensors (11 sensors), kitchen-mounted sensors (9 sensors), and bedroom/bathroom sensors (10 sensors). The sensors within each group were concurrently activated and synchronised using the Xsens DOT applications installed on two tablets and a smartphone which communicated with sensors via Bluetooth. Subsequently, synchronisation across the three sensor groups was achieved by temporally aligning sensor timestamps in the recorded data with the manually annotated activity labels.

    Sensors were turned on at the commencement of each experimental session and remained operational until the completion of the final repetition to ensure continuous data capture. Following the conclusion of the experiments, the recorded data were downloaded from the sensors using the DOT Data Exporter software via a wired connection to ensure reliable and lossless transfer. All repetitions performed during a session were stored collectively as a single continuous data file.

    \subsection{Data pre-processing}
    
        After data collection was completed, each continuous sensor dataset underwent a structured preprocessing stage in which the data were segmented into individual repetitions based on temporal markers. These repetitions were then further partitioned into distinct activity segments using the corresponding activity labels collected during the experiments. This multi-stage segmentation process facilitated the creation of well-defined activity-specific datasets suitable for subsequent analysis and HAR model development.
        To verify the accuracy of the segmentation, the activity-wise data from each participant were visualised and systematically inspected. This verification step ensured that the temporal boundaries of each segment are aligned with the annotated activity labels and that the recorded sensor signals appropriately reflected the expected motion characteristics of the intended activities.

        During this process, minor instances of data leakage were identified, where samples from one activity class were inadvertently included in another. For example, a small number of samples associated with the sitting activity were found to overlap with segments labelled as standing. If left uncorrected, such inconsistencies could negatively affect model performance and overall data reliability. Following manual inspection, mislabelled samples were either corrected or removed from the dataset. Samples were relabelled when the true activity could be clearly identified from the available sensor data and contextual information. In cases where the correct label could not be established with sufficient confidence, the samples were removed to avoid introducing uncertainty or bias into the dataset. This quality assurance process improved data integrity and strengthened the reliability of downstream activity recognition analyses.

\section{Data Record}\label{sec4}

\begin{figure}[b!]
\centering
\includegraphics[width=0.9\linewidth]{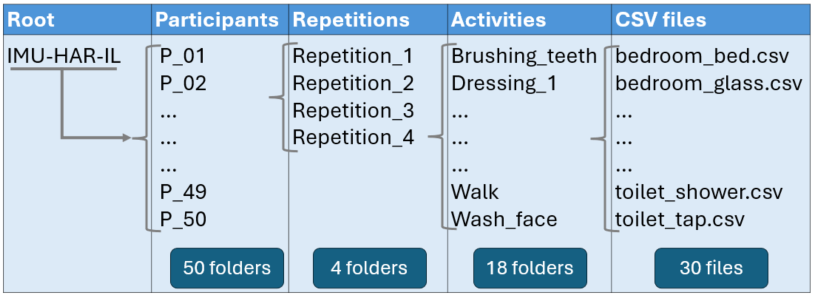}
\caption{Data organisation in structured folders.}
\label{fig:data_organisation}
\end{figure}

    The dataset is organised in a hierarchical directory structure to ensure systematic storage, traceability, and ease of access for analysis (Fig.~\ref{fig:data_organisation}). At the top level, HAR-IMU-IL data are stored as comma-separated values (CSV) files distributed across 50 primary folders, each corresponding to an individual participant. This participant-centric organisation facilitates subject-level analysis and supports reproducibility by clearly separating data collected from different individuals. Within each participant folder, up to four subfolders are included to represent multiple data collection sessions or repetitions, reflecting repeated sessions conducted under similar experimental conditions.

    Each repetition subfolder is further subdivided into 18 activity-specific folders (the dressing activity was performed twice in each repetition, as shown in Table~\ref{table:activity_interpretation}), corresponding to the distinct activities performed by the participant during data collection. This activity-based segregation enables efficient retrieval and comparison of sensor data across different contexts. Within each activity folder, data are stored for the 30 individual IMU sensors, with each sensor’s recordings maintained as separate CSV files. This fine-grained organisation preserves the integrity of multiple sensor streams while allowing flexible aggregation and preprocessing at the sensor, activity, repetition, or participant level, thereby supporting a wide range of analytical and machine learning workflows. Table~\ref{table:size_dataset} describes the total amount of data collected during the various activities and in different functional areas.

    \begin{table}[b!] 
    \caption{Overall activity duration in the dataset.}\label{table:size_dataset}
    \centering
    \begin{tabular}{p{0.20\linewidth} p{0.23\linewidth} p{0.20\linewidth} p{0.23\linewidth}}
    \toprule
    \textbf{Functional area}	& \textbf{Activity} 	& \textbf{Duration (s)} 	& \textbf{Total duration (s)} \\
    \midrule
    \multirow{8}{*}{Mobility} &	Sitting & 17560 & \multirow{8}{*}{113353}\\
     & Standing & 17715  &\\
     & Walking & 20932 & \\
     & Lying & 19707  &\\
     & Stairs & 16797  &\\
     & Sit-stand transitions & 10183 & \\
     & Sit-lay transitions & 10459 & \\
    \hline
    \multirow{5}{*}{Hygiene} & Washing face & 4672 & \multirow{5}{*}{42183}\\
     & Brushing teeth & 5930 & \\
      & Using toilet & 8007 & \\
       & Showering & 9869 & \\
        & Dressing & 13705 \\
    \hline
    \multirow{4}{*}{Nutr. \& hydr.} & Meal preparation & 17520 & \multirow{4}{*}{38277} \\
    & Drinking & 3497 & \\
    & Eating & 15824 & \\ 
    & Empty kitchen bin & 1436 & \\
    \hline
    Medication & Taking medicine & 5297 & 5297 \\
    \bottomrule
    \end{tabular}
    \end{table}

    \subsection{Data description}
    The data recorded by the IMU sensors is stored in CSV files with multiple columns. These files contain the following types of data:
    \begin{itemize}
        \item \texttt{Euler\_X}, \texttt{Euler\_Y}, \texttt{Euler\_Z}: The representation of 3D orientation using three sequential angles: Roll, Pitch, and Yaw.
        \item \texttt{Acc\_X}, \texttt{Acc\_Y}, \texttt{Acc\_Z}: Acceleration along x, y, and z directions.
        \item \texttt{Gyr\_X}, \texttt{Gyr\_Y}, \texttt{Gyr\_Z}: The representation of the rate of turn (angular velocity).
        \item \texttt{Activity\_label}: A unique number to identify the activity.
    \end{itemize}

    \subsection{Metadata}
    The \textit{README.txt} file stored on the data portal contains information about: (i) the participants, (ii) the IMU sensor and the sampling rate, (iii) the number of deployed sensors, (iv) the number of considered activities, (v) the distribution of data folders, (vi) the CSV file format, (vii) the activity labels, and (viii) the data collection duration.


\begin{figure*}[b!]
\centering
\includegraphics[width=0.9\linewidth]{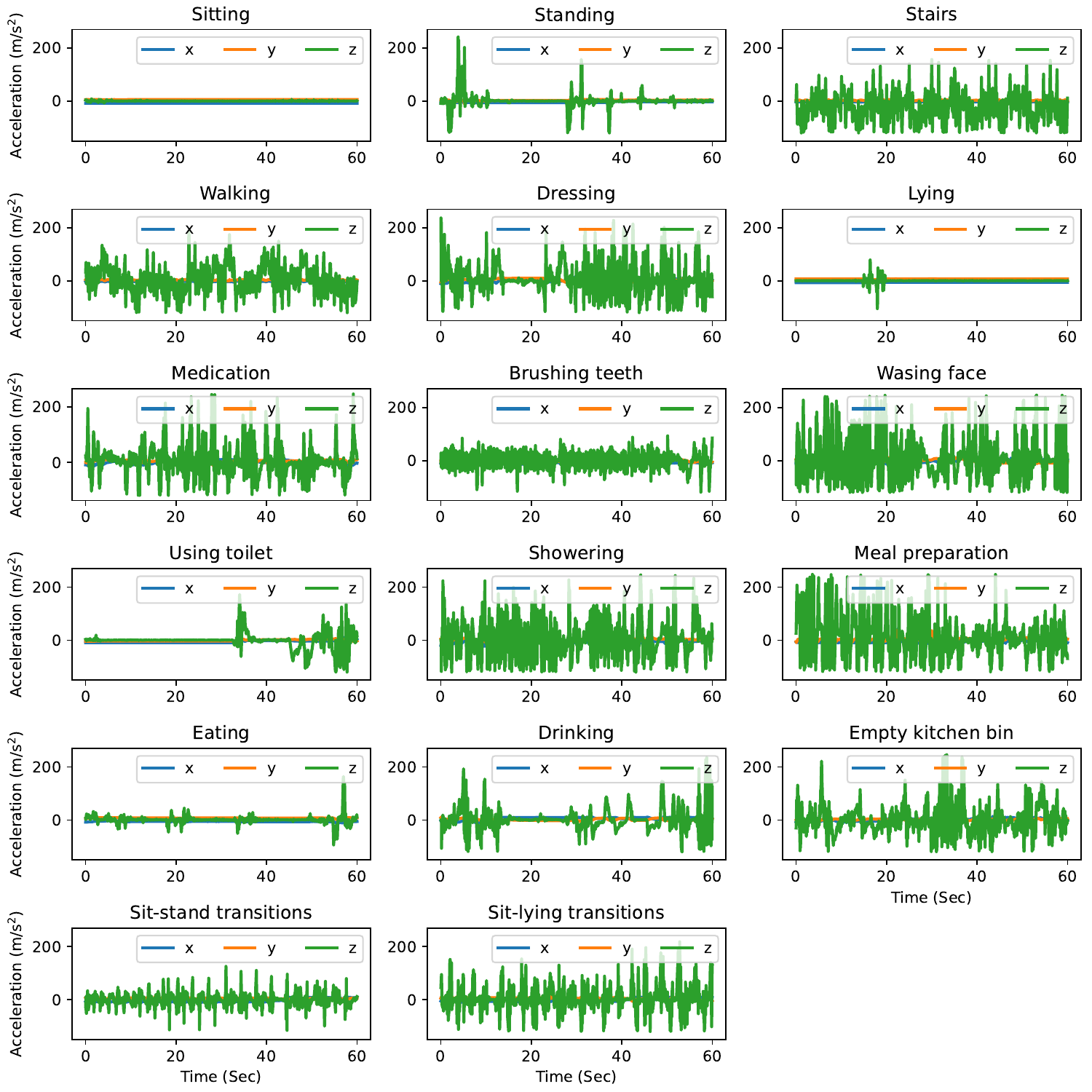}
\caption{3-axis accelerometer signals from a wrist-worn IMU sensor exhibit distinct patterns during various physical human activities.}
\label{fig:accelerometer_signal}
\end{figure*}

    \begin{figure}[htbp]
      \centering
      \begin{subfigure}[b]{0.7\textwidth}
        \centering
        \includegraphics[width=\textwidth]{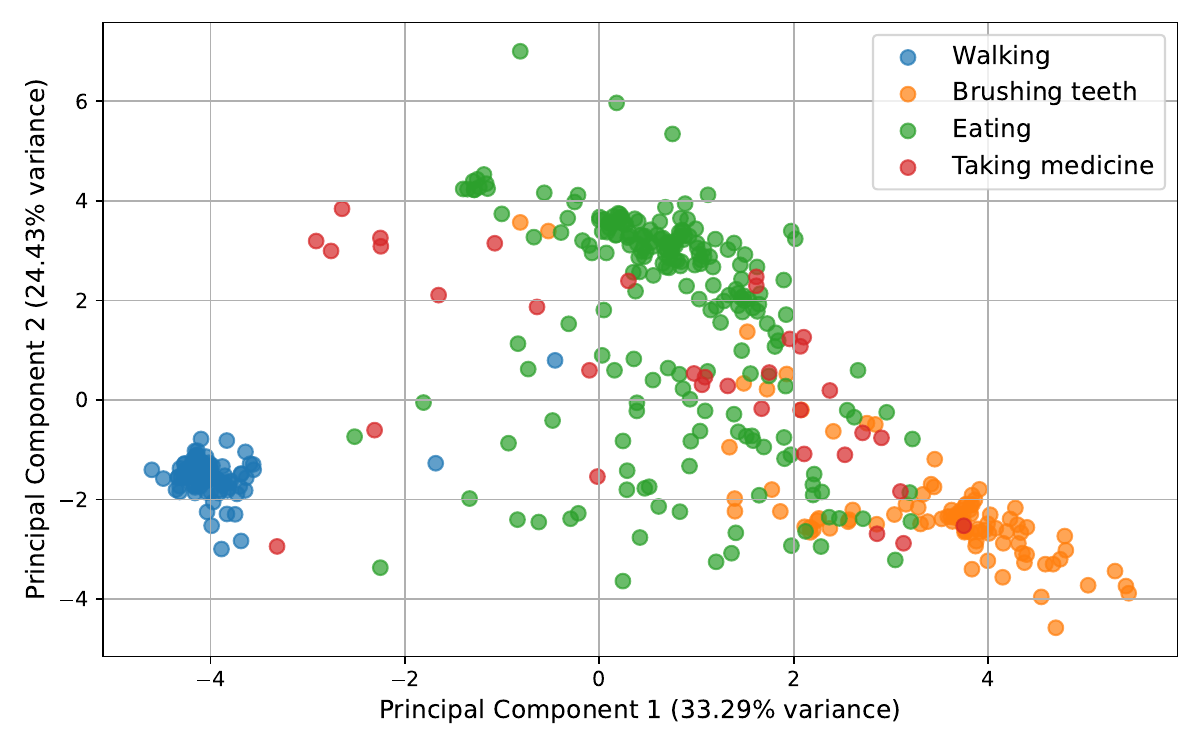}
        \caption{Participant 1}
        \label{fig:sub1}
      \end{subfigure}
      
      \begin{subfigure}[b]{0.7\textwidth}
        \centering
        \includegraphics[width=\textwidth]{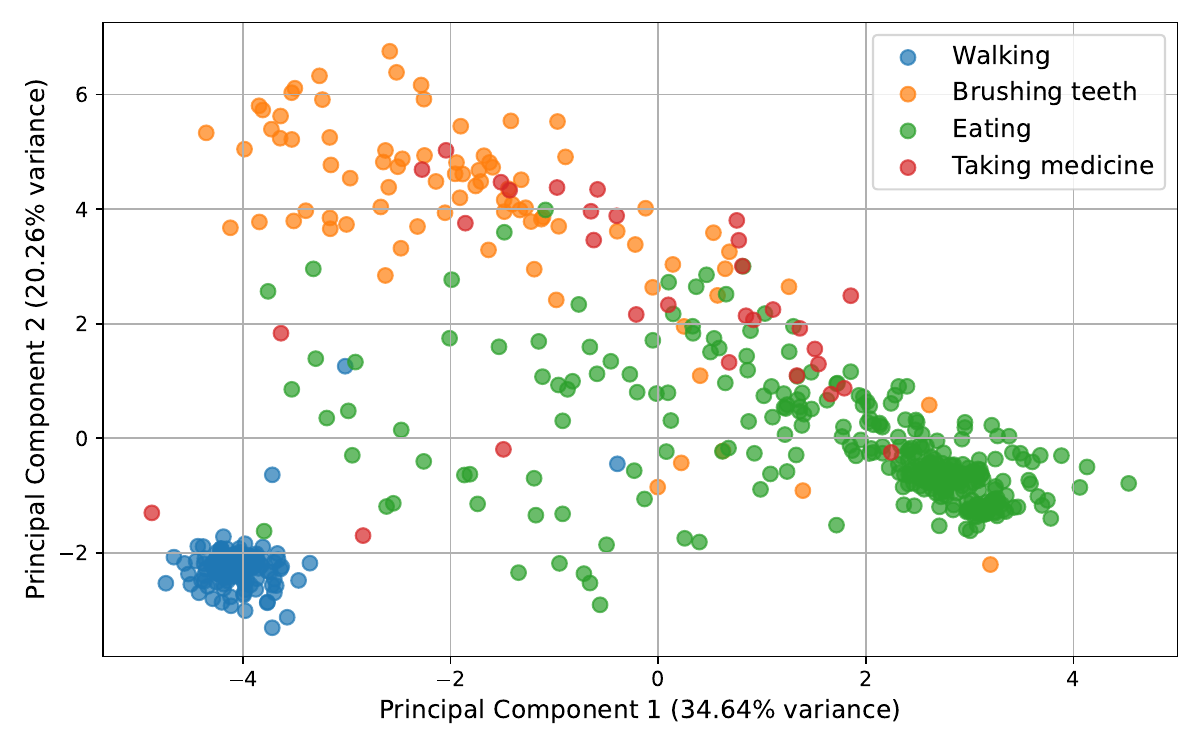}
        \caption{Participant 2}
        \label{fig:sub2}
      \end{subfigure}

        \begin{subfigure}[b]{0.7\textwidth}
        \centering
        \includegraphics[width=\textwidth]{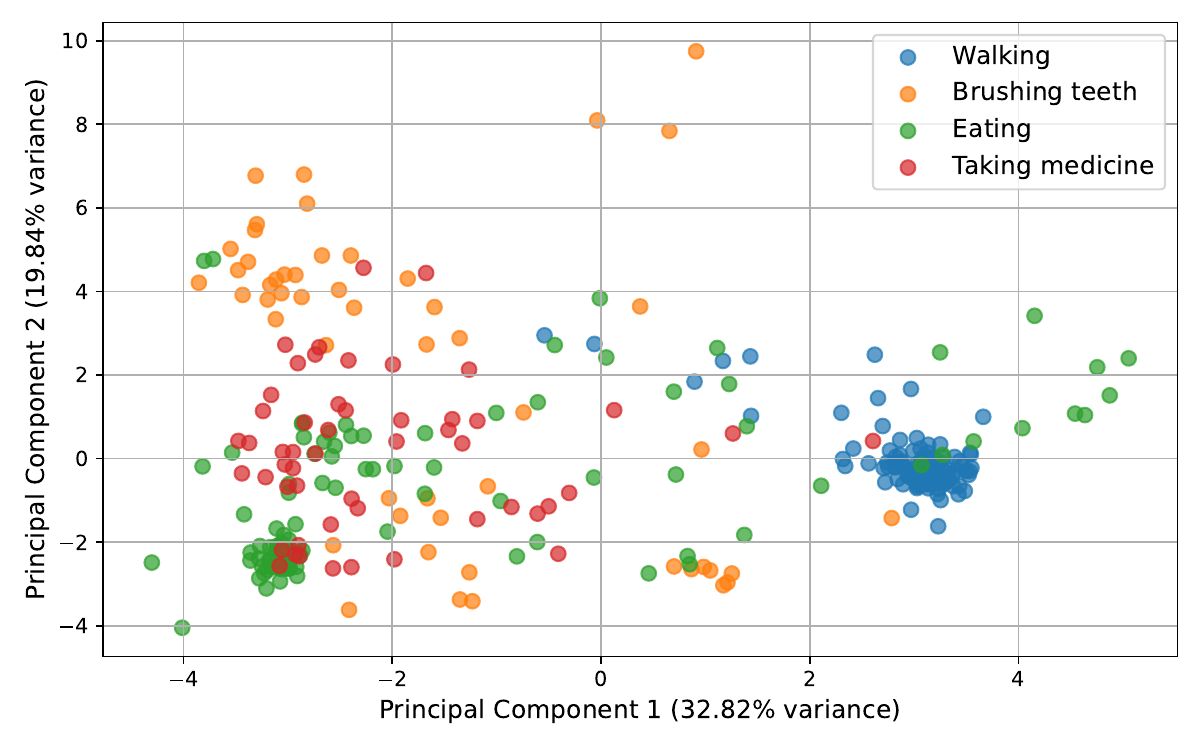}
        \caption{Participant 3}
        \label{fig:sub3}
      \end{subfigure}
      
      \caption{PCA analysis of 3-axis accelerometer signals from the (right) wrist-worn sensor.}
      \label{fig:PCA_analysis}
    \end{figure}

\section{Technical Validation}

    This section provides an empirical assessment of the quality, structure, and discriminative potential of the HAR-IMU-IL dataset. We first examine representative sensor signals to highlight activity-specific motion patterns, followed by an analysis of feature separability using dimensionality reduction. Finally, we evaluate the dataset’s effectiveness for HAR through a baseline machine learning pipeline and report results in terms of activity recognition accuracy.

    Representative \SI{60}{\second} accelerometer signal traces corresponding to multiple activities captured using a wrist-worn IMU are presented in Fig.~\ref{fig:accelerometer_signal}. Clear differences in both temporal structure and signal amplitude are observed across activities, revealing distinct motion signatures. These characteristic patterns highlight the discriminative capacity of IMU data and emphasise their suitability for reliable activity recognition.

    To further investigate the underlying structure and variability of the sensor data, principal component analysis (PCA) was performed on the three-axis accelerometer signals collected from the wrist-worn sensor. Fig. \ref{fig:PCA_analysis} presents the PCA projections for three different participants performing four representative activities corresponding to four distinct functional areas. The PCA visualisation reveals substantial overlap among the activities of brushing teeth, eating, and taking medicine across all participants. This overlap can be attributed to the similarity of upper-limb movement patterns involved in these activities, as they primarily consist of repetitive hand-to-mouth or hand-to-face motions performed within a limited range of motion. As a result, the extracted accelerometer features exhibit comparable characteristics, resulting in reduced separability in the lower-dimensional PCA space.
    
    In contrast, the walking activity forms a relatively distinct cluster that is clearly separated from the other activities. This separation is likely due to the unique rhythmic and whole-body movement patterns associated with locomotion, which generate characteristic acceleration signatures that differ substantially from those observed during the other activities. The consistency of this clustering across participants suggests that walking exhibits stronger discriminative features and lower inter-class similarity compared to the remaining activities.

    Following these exploratory analyses, the dataset was used to develop and evaluate baseline machine learning models for activity recognition. To promote robust model generalisation and minimise overfitting, the dataset was divided into training and testing subsets using an 80/20 holdout strategy. Specifically, 80\% of the data were allocated to model training and internal validation. The remaining 20\% of the data were reserved as an independent test set, providing an unbiased assessment of the finalised models on previously unseen data.

    \begin{figure}[b!]
    \centering
    \includegraphics[width=\linewidth]{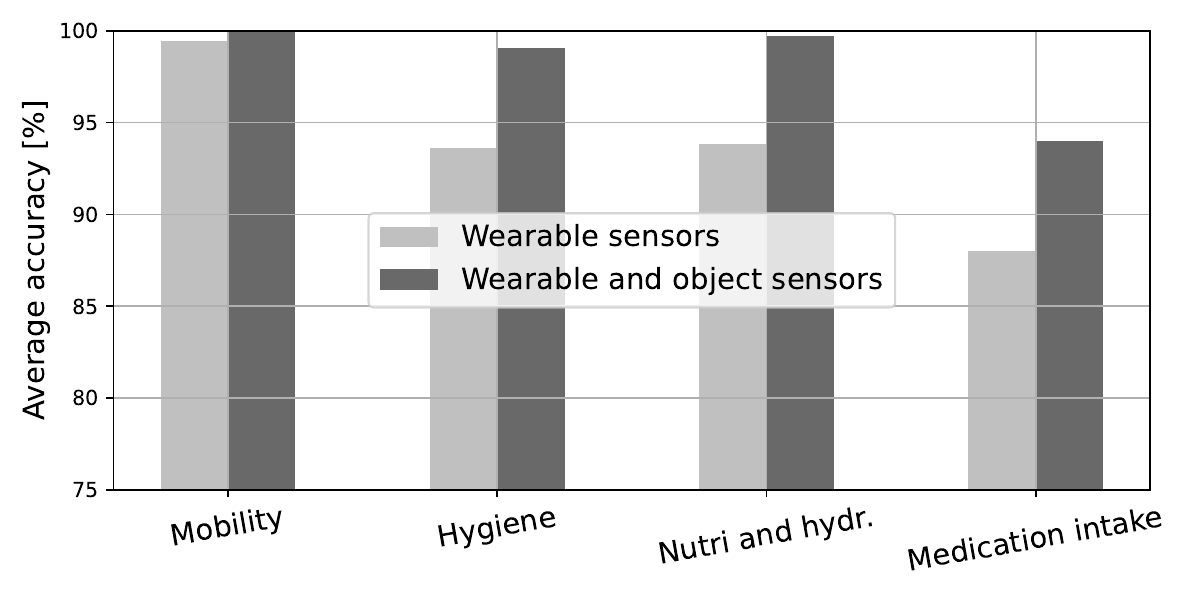}
    \caption{Performance of RF model to recognise 4 functional areas using wearables sensors, and fusion of wearable and object-mounted sensors (average accuracy: 97.19\% using wearables; 99.58\% using fusion of wearable and object-mounted sensors).}
    \label{fig:barplot_functional_area}
    \end{figure}

    \begin{figure}[b!]
    \centering
    \includegraphics[width=\linewidth]{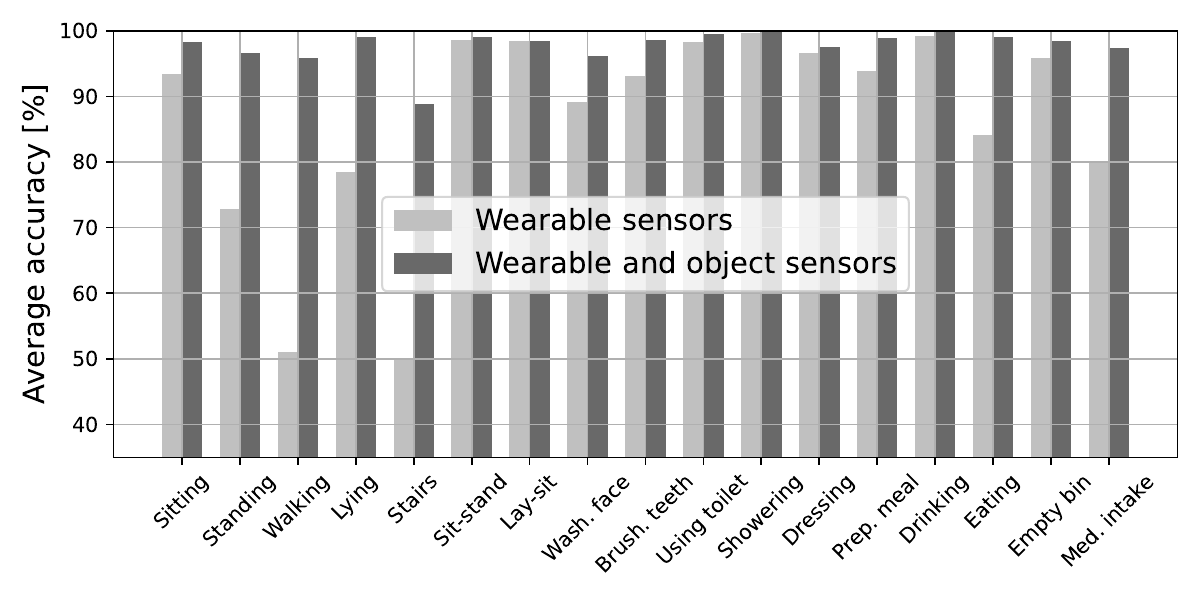}
    \caption{Performance of RF model to recognise 17 activities using wearables sensors, and fusion of wearable sensors and object-mounted sensors (average accuracy: 92.83\% using wearables; 98.66\% using fusion of wearable and object-mounted sensors).}
    \label{fig:barplot_activities}
    \end{figure}
    
    Following data partitioning, the time-series signals were segmented into 1-second sliding windows with a 50\% overlap to capture both temporal continuity and sufficient activity context. A comprehensive set of time- and frequency-domain features was then extracted from each window~\cite{sandhu2023fusedar,sandhu2021solar,sandhu2020towards_energy_positive}. These features served as inputs to Random Forest (RF) classifiers developed for the recognition of functional areas and human activities. The selected feature set was designed to capture both motion dynamics and contextual information derived from the deployed sensing system.
    As shown in Fig.~\ref{fig:barplot_functional_area}, the RF model achieved an average functional area recognition accuracy of 97.19\% using wearable sensors. By combining wearable and object-mounted sensor data, the accuracy increased to 99.58\%, highlighting the effectiveness of multisensor fusion in enhancing activity recognition performance.
    Similarly, Fig.~\ref{fig:barplot_activities} shows strong performance for activity recognition across a diverse set of activities. A consistent trend observed in both figures is the improvement in classification performance when object-mounted sensor data are fused with wearable sensor data, compared to using wearable sensors alone. This finding highlights the value of incorporating environmental context into HAR systems, enabling the model to better distinguish between activities that may exhibit similar body movement patterns.
    
    A comparison of the two classification tasks further reveals that functional area recognition achieves higher accuracy than activity recognition. This difference is likely due to the greater complexity of activity classification, which involves a larger number of classes and higher intra-class variability. The observed performance gains from sensor fusion can be attributed to the complementary information provided by the wearable and object sensors. While wearable sensors primarily capture user motion and posture, object-mounted sensors provide contextual cues related to object usage and environmental interactions. The combination of these perspectives yields richer and more discriminative feature representations, leading to enhanced recognition performance. Additional analyses and detailed experimental evaluations on a subsample of the dataset are reported in~\cite{sandhu2025feasibility,sandhu2025fusing}.
    
\section{Usage Notes}

    This paper introduced HAR-IMU-IL, a comprehensive HAR dataset collected in real‑world home environments using wearable and object‑mounted IMU sensors. The dataset captures 17 key physical activities across four functional domains that are central to assessing independent living capabilities of older adults including mobility, hygiene, nutrition and hydration, and medication intake. By leveraging a dense deployment of 30 off‑the‑shelf IMU sensors, the dataset provides realistic and high‑quality motion data suitable for developing and evaluating HAR models in smart home settings. Detailed descriptions of the data collection protocol, annotation process, and baseline performance benchmarks are provided to support reproducibility and facilitate future research on activity recognition, functional monitoring, assistive technologies, and smart home automation. While the dataset provides high-quality, multi-sensor recordings with detailed annotations, several practical considerations should be taken into account when using it.

    Some segments of the dataset were affected by limitations associated with practical and environmental constraints encountered during data acquisition. For example, a small number of stair-related activities could not be recorded due to unfavourable outdoor weather conditions, such as rainfall. In addition, certain recordings were impacted by sensor malfunction or temporary sensor misalignment during data collection. Consequently, a limited number of experimental trials contain missing data from one or more IMU sensors. Although extensive care was taken to preprocess, synchronise, and align all sensor streams to ensure data quality and consistency, residual imperfections remain unavoidable. In particular, minor temporal overlap or leakage between adjacent activities may persist, which should be carefully considered when developing, evaluating, and interpreting predictive models based on this dataset.

    Although the dataset was collected from 50 participants using 30 IMU sensing devices, variations in data completeness and activity execution were observed across participants. In particular, some participants completed fewer than four repetitions of all or certain physical activities, resulting in an unequal number of samples across activity classes. Furthermore, most participants performed the dressing activity while standing, whereas a small number completed the activity while seated on the bed. Similarly, the medication intake activity was predominantly performed while sitting on the bed, although a minority of participants performed it in a standing position. The drinking activity was also mainly conducted while sitting, with only a few participants performing it while standing. In several cases, participants washed plates immediately after eating, and this behaviour was considered part of the eating activity.
    
    It is important to note that certain parts of activities in the dataset were simulated (Table \ref{table:activity_interpretation}), which represents a limitation of the study. For example, activities such as brushing teeth, showering, using toilet, preparing meal, and emptying kitchen bin tasks were performed according to predefined instructions, rather than being captured during the participants’ routine daily lives. Consequently, these activities may not fully reflect real-world behavioural patterns. Therefore, these considerations must be kept in mind while developing, testing and interpreting the activity recognition models using this dataset.

    \section{Data Availability}
    The HAR-IMU-IL dataset is publicly available online on CSIRO's Data Access Portal at the following link:
    \url{https://doi.org/10.25919/d7xf-n080}.
    
    The public repository includes the CSV data files along with a \textit{README.txt} file describing the data format and other necessary information.

    \section{Code Availability}

    The codebase for reading, preprocessing, and implementing the baseline machine learning algorithms reported in this study is publicly available on GitHub at: \url{https://github.com/mmsandhu/IMU-HAR-IL}. The implementation was developed using Python 3.9 and is provided in \textit{.py} format (\url{https://www.python.org/}). Specifically, the Pandas library (\url{https://pandas.pydata.org/}) was used for loading CSV files, the NumPy library (\url{https://numpy.org/}) was employed for data preprocessing, segmentation, and feature extraction, and the Scikit-learn library (\url{https://scikit-learn.org/}) was utilised for training and evaluating the baseline machine learning model.

\backmatter

\bmhead{Author contributions}

M.S. was responsible for study design, collecting, curating, and analysing the data, and took the lead in writing the manuscript. W.L. contributed in study design and data collection. B.K. assisted with the development of the study design and data collection. D.S.T. contributed in study design and in structuring and refining the manuscript. All authors reviewed the manuscript.

\bmhead{Competing interests}

The authors declare no competing interests.

\bmhead{Acknowledgements}

The authors would like to acknowledge the strategic investment of the Future Science Platform program at CSIRO and extend their thanks to the clinicians and participants who contributed to this study.

\bmhead{Funding}
No external funding was received for conducting this study.

\bmhead{Ethics statement}
Ethics approval to complete this study was obtained from CSIRO's Health and Medical Human Research Ethics Committee, before conducting this study (Approval number: 2024\_018\_LR). All participants provided informed written consent before participating in the study.

\bibliography{bibliography}

\end{document}